\documentclass[runningheads]{svmult}

\usepackage{makeidx}   
\usepackage{graphicx}  
\usepackage{subeqnar}  
\usepackage{multicol}  
\usepackage{physprbb}  
\makeindex             

\begin{document}
\title*{Stability of a Vortex in a Rotating Trapped Bose-Einstein
Condensate\thanks{This work  was
supported in part by NSF Grant No.~99-71518.  The manuscript benefited from
our participation in workshops organized by H.~Stoof at  the Lorentz Center,
Leiden, The Netherlands and by S.~Stringari at ECT$^*$, Trento, Italy; we are
grateful  for their hospitality.}}
\toctitle{Stability of a Vortex in a Rotating
\protect\newline  Trapped Bose-Einstein Condensate}
%
%
\titlerunning{Stability of a Vortex}
%
\author{Alexander L.~Fetter
\and Anatoly A.~Svidzinsky\thanks{Supported in part by Stanford
University.}}

\institute{Geballe Laboratory for Advanced Materials, Stanford University,
Stanford, CA 94305-4045, USA}

\maketitle              


\section{Time-Dependent Gross-Pitaevskii Equation}

The remarkable achievement of Bose-Einstein condensation in dilute
trapped alkali-metal atomic gases~\cite{Ande95,Brad95,Davi95} has stimulated
the (now successful) search for quantized vortices that are usually
associated with  external
rotation~\cite{Matt99,Madi99,Madi00,Chev00,Ande00}.  An essential feature of
these condensates is the
  order parameter, characterized by a complex macroscopic wave
function $\Psi({\vec r},t)$.  Theoretical descriptions of
vortices in  trapped low-temperature condensates have relied on the
time-dependent Gross-Pitaevskii (GP) equation~\cite{Gros61,Pita61,Dalf99},
which omits dissipation.  For a  trap rotating at
angular velocity
$\Omega$ about $\hat z$, it is  a nonlinear Schr\"odinger
equation

\begin{equation}
i\hbar\frac{\partial\Psi}{\partial t}= \left(T+V_{\rm
tr}-\Omega\,L_z+g|\Psi|^2\right)\Psi,\label{GP}
\end{equation}
where $T=-\hbar^2 \nabla^2/2M$ is the kinetic-energy operator, $V_{\rm
tr}=\frac{1}{2}M\sum_{j}\omega_j^2x_j^2$ is the harmonic trap potential, $L_z
= xp_y-yp_x$ is the $z$ component of angular momentum, and
$g\equiv 4\pi \hbar^2 a/M$ characterizes the strength of the short-range
interparticle potential (here, $a$ is the positive  $s$-wave
scattering length for repulsive two-body interactions; typically $a $ is a
few nm). Current experiments involve ``dilute'' systems, so that the
zero-temperature condensate contains nearly all $N$ particles, with  $\int
dV\,|\Psi|^2
\approx N$. For a steady solution,
$\Psi(\vec{r},t)=\Psi(\vec{r})e^{-i\mu t/\hbar}$, where $\mu$ is the chemical
potential.

\subsection{Equivalent Hydrodynamics of Compressible Isentropic Fluid}

If the condensate wave function is written as $\Psi = e^{iS}|\Psi|$,
the real and imaginary parts of Eq.~(\ref{GP}) precisely reproduce the
time-dependent  irrotational  hydrodynamics of a compressible isentropic
fluid, written in terms of the   number density
$n=|\Psi|^2$ and the velocity potential $\Phi = \hbar S/M$, where $\vec{v}=
\vec{\nabla}\Phi$~\cite{Gros61,Pita61,Fett96}.  In principle, the dynamics
of a curved vortex line in a rotating trap follows  directly
from general hydrodynamics, but the trap
potential and resulting nonuniform density  complicate the problem
considerably. Thus it is preferable  to  start with simple
situations.

\subsection{Thomas-Fermi Limit for Large Condensates}
For a noninteracting condensate, the macroscopic wave function is the
ground state of the harmonic oscillator, with spatial extent
$d_j=(\hbar/M\omega_j)^{1/2}$ and $j = x,y,z$; typically $d_j$ is a few
$\mu$m.  The repulsive interactions act to expand the condensate, and  the
relevant dimensionless interaction parameter is $Na/d_0$, where
$d_0=\left(d_xd_yd_z\right)^{1/3}$ is a suitable geometric mean.  Recent
experiments focus on the regime
$Na/d_0\gg 1$, when  the quantum-mechanical energy associated with  the
density gradients is negligible.  In this Thomas-Fermi (TF)
limit~\cite{Baym96}, the nonrotating GP equation has the solution
$g|\Psi_{TF}|^2+V_{\rm tr} \approx \mu$, with a    parabolic density profile
\begin{equation}
n_{TF}(\vec{r}) \approx
n_{TF}(0)\left(1-\frac{x^2}{R_x^2}-\frac{y^2}{R_y^2}
-\frac{z^2}{R_z^2}\right).\label{TF}
\end{equation}
Here
$R_j =
(2\mu/M\omega_j^2)^{1/2}\gg d_j$ fixes the condensate's dimensions and
$n_{TF}(0) =
\mu/g$ is the central density.  The (large) dimensionless ratios
$R_0^5/d_0^5 = 15 Na/d_0$ and $\mu/\hbar\omega_0 = \frac{1}{2}R_0^2/d_0^2$
characterize the effect of the repulsive interactions, where $\omega_0$ and
$R_0$ are  appropriate geometric means.

\section{Energy of a Vortex in a Large Rotating Trap}

The time-dependent GP equation (\ref{GP}) follows from a variational
energy functional

\begin{equation}
E(\Omega) = \int dV\,\Psi^*\left(T+V_{\rm
tr}+\textstyle{\frac{1}{2}}g|\Psi|^2-\Omega\,L_z\right)\Psi\,.\label{en}
\end{equation}
If $\Psi$  represents a straight singly quantized vortex in a
  rotating disk-shape trap ($R_\perp\gg R_z$), the TF density
  yields
  the increase in energy
$\Delta E(\vec{r}_0,\Omega)$ associated with the vortex at the
transverse position~$\vec{r}_0$~\cite{Svid00}. For a nonrotating trap with
$\Omega=0$, this energy $\Delta E(\vec{r}_0,0)$ decreases monotonically with
lateral displacement.  Energy conservation requires that the allowed motion
is a precession at fixed trap potential (the trajectory is elliptical for an
anisotropic trap).  In the presence of weak dissipation, the vortex slowly
spirals outward, lowering its energy.

For arbitrary $\Omega$ and small lateral
displacements, $\Delta E(\vec{r}_0,\Omega) $ has
the form

\begin{eqnarray}
\Delta E(\vec{r}_0,\Omega)\ \approx&\displaystyle \frac{4\pi}{3}\,\frac{R_z
n(0)
\hbar^2}{M}\bigg\{
\left[\ln\left(\frac{R_\perp}{\xi}\right)-\frac{2 M
\Omega R_\perp^2}{5 \hbar }\right]\nonumber\\ \noalign{\vskip.1cm}
&\displaystyle
-\frac{1}{2}\left(\frac{x_0^2}{R_x^2}+\frac{y_0^2}{R_y^2}\right)
\left[3\ln\left(\frac{R_\perp}{\xi}\right)-\frac{2 M
\Omega R_\perp^2}{\hbar}\right]\bigg\},\label{approx}
\end{eqnarray}
where $\xi =\hbar/(2M\mu)^{1/2}$ is the vortex-core radius and $2/R_\perp^2
= 1/R_x^2+1/R_y^2$ defines the mean transverse radius of the condensate. In
the TF limit,  $\xi$ is small, with $\xi R_0 =d_0^2$,
  ensuring a clear separation of length scales $\xi\ll d_0\ll R_0$.

\begin{itemize}

\item For small positive $\Omega$, $\Delta E(\vec{r}_0,\Omega)$ decreases
with increasing  lateral displacements so that  a vortex  will spiral
outward in the presence of weak dissipation.

\item  The  curvature at $r_0=0$ vanishes at $\Omega_m =
\frac{3}{2}(\hbar/MR_\perp^2)\ln(R_\perp/\xi)$, signaling the onset of
metastability. Since $\Delta E(0,\Omega_m)>0$, this
state  is not truly stable.

\item  When $\Omega>\Omega_m$,  the trap center becomes a local {\em
minimum}.  For weak dissipation, a vortex that is slightly  displaced from
the center  will spiral {\em inward}.

\item When $\Omega$ reaches $\Omega_c =
\frac{5}{2}(\hbar/MR_\perp^2)\ln(R_\perp/\xi)= \frac{5}{3}\Omega_m$, a
central vortex becomes stable because $\Delta E(0,\Omega_c)=0$.  Hence
$\Omega_c$ is the thermodynamic critical angular velocity for vortex
creation.

\item For $\Omega>\Omega_c$, the form of $\Delta E(\vec{r}_0,\Omega)$
shows that a vortex at the outer edge remains metastable, but the barrier for
entry into the condensate decreases.    Spontaneous nucleation of a
vortex may eventually occur through a surface
instability~\cite{Dalf97,Fede99,Isos99}.   The observed critical angular
velocity for vortex creation~\cite{Madi99,Chev00} is $\approx 70\%$ higher
than the predicted TF value
$\Omega_c$, in qualitative agreement with such a surface mechanism (but see
Ref.~\cite{Fede00} and Sec.~4.2 for an alternative explanation).

\end{itemize}

\section{Small-Amplitude Excitation of a Vortex in a  Rotating Trap}

A macroscopic  Bose condensate acts like an external
  particle source, in the sense that  the same  physical excited state
(here labeled by $j$)  can be achieved either by adding or by subtracting one
particle.  The actual eigenstates  involve ``quasiparticle''
operators
$\alpha_j^\dagger$ and $\alpha_j$ that are linear combinations of the two
quantum-mechanical states.  For a given normal mode, the resulting pair of
coupled complex amplitudes $u_j(\vec{r})$ and $v_j(\vec{r})$ obey  the
Bogoliubov equations~\cite{Pita61,Fett72} that determine  the corresponding
eigenfrequency $\omega_j$.  Imposing Bose-Einstein commutation relations
$[\alpha_j,\alpha_k^\dagger] = \delta_{jk}$ requires that
these amplitudes obey the particular normalization
$\int dV\,(|u_j|^2-|v_j|^2)=1$, and the resulting quasiparticle
Hamiltonian reduces to a set of uncoupled harmonic oscillators
$\sum_j\hbar\omega_j\alpha_j^\dagger\alpha_j$, summed over all modes with
positive normalization.  In the simplest case of a uniform condensate, the
  eigenfrequencies are all positive, which ensures that the system is
stable because the energy then has a lower bound. For a uniform  condensate
moving with velocity
$\vec v$,  however, some of the eigenfrequencies can become negative as soon
as
$|\vec{v}|$ exceeds the speed of  sound, which corresponds to the
instability associated with the Landau critical velocity~\cite{Land41}.
Physically, the system can spontaneously generate quasiparticles
because the Hamiltonian is no longer bounded from below.

These general ideas have direct relevance to the stability of a singly
quantized vortex in a rotating trapped condensate.  For simplicity, it is
convenient to consider an axisymmetric condensate in equilibrium with a
  trap rotating at  an angular velocity $\Omega$.  In this case, states can
be characterized by their azimuthal angular quantum numbers $m_j$, and the
transformation to rotating coordinates
$H\to H-\Omega L_z$ ensures that the eigenfrequencies $\omega_j(\Omega)$ in
the rotating frame are simply $\omega_j(\Omega)= \omega_j-m_j\Omega$, where
$\omega_j$ is the corresponding eigenfrequency in the nonrotating frame.

\subsection{Stability of a Vortex}

The first numerical study of the Bogoliubov equations for a singly quantized
vortex in a nonrotating trap found a single negative-frequency mode
(the ``anomalous mode'')~\cite{Dodd97,Rokh97}, implying that the vortex in
the condensate is unstable.  Physically, this instability arises because the
condensate with a vortex has a higher energy than the vortex-free
condensate.

This situation is especially clear for  a noninteracting Bose gas in an
axisymmetric harmonic trap, when the first excited (vortex) state has an
excitation energy
$\hbar\omega_\perp$ and an angular momentum  $\hbar$.  The
transition back to the true ground state involves the (negative) frequency
$-\omega_\perp$ and the (negative) change in angular quantum number $-1$.
More generally, the numerical analysis for small and medium interaction
strength $N a/d_\perp$~\cite{Dodd97} found that the anomalous
frequency
$\omega_a$ remained negative (with $m_a=-1$  unchanged).  In
a frame rotating with angular velocity $\Omega$, the
  anomalous frequency  becomes
$\omega_a(\Omega) = \omega_a+\Omega.\label{anom}
$
Since $\omega_a<0$, the eigenfrequency
$\omega_a(\Omega)$ in the rotating frame rises toward
$0$ with increasing
$\Omega$ and vanishes at a critical rotation speed $\Omega^* = -\omega_a =
|\omega_a|$.  The Bogoliubov description of small oscillations
implies that a condensate with a singly quantized vortex is {\em unstable}
for
$\Omega<\Omega^*$ but becomes {\em stable} for
$\Omega>\Omega^*$.

In contrast to the numerical study~\cite{Dodd97} for small and medium values
of the dimensionless  coupling parameter
$Na/d_0$, a direct perturbation analysis is feasible for a large
condensate ($Na/d_0\gg 1$)  containing an axisymmetric singly quantized
vortex (the TF limit). Detailed study of the Bogoliubov
equations~\cite{Svid00} yields the explicit expression for the anomalous
frequency in the rotating frame
$\omega_a(\Omega) =
-\frac{3}{2}\left(\hbar/M
R_\perp^2\right)\ln\left(R_\perp/\xi\right)+\Omega$, which has  the
expected form.  The resulting critical angular velocity
$\Omega^* = \frac{3}{2}\left(\hbar/M
R_\perp^2\right)\ln\left(R_\perp/\xi\right)$ for the onset of local
stability agrees  with
$\Omega_m$ inferred from Eq.~(\ref{approx}) for the onset of metastability.

\subsection{ Splitting of Normal-Mode Frequencies Caused by a Vortex}

The ground-state condensate can sustain dynamical oscillations driven by the
mean-field repulsive interaction (analogous to plasma oscillations in a
charged medium).  These normal modes become particularly simple in the TF
limit of a large condensate~\cite{Stri96}, and experiments have confirmed
the predictions for the lowest few modes in considerable
detail~\cite{Dalf99}.  For an axisymmetric condensate, the normal modes can
be classified by their azimuthal quantum number $m$, and modes with $\pm m$
are degenerate.

	When the condensate contains a vortex, the asymmetric circulating flow
affects the preceding normal modes. In particular, the
originally degenerate modes are split  by the Doppler shift of the local
frequency (analogous to the splitting of
magnetic sublevels in the Zeeman effect).  In the TF limit, this 
small fractional splitting of the
degenerate modes is proportional to $|m|d_0^2/R_0^2$~\cite{Svid98,Zamb98};
it has served to detect the presence of a
vortex and to infer the circulation and angular
momentum~\cite{Chev00,Corn00}.

\section{Vortex Dynamics}
At zero temperature, the time-dependent GP equation (\ref{GP}) determines
the dynamics of the condensate in a rotating trap.  A vortex line will move
in response to the nonuniform trap potential, the external rotation,
and its own  local curvature.  This problem is
especially tractable in the TF limit, because the small vortex core radius
$\xi$ permits a clear separation of length scales; the method
of matched asymptotic expansions yields an explicit
expression for the local velocity of a vortex line~\cite{Pism91,Rubi94}.

\subsection{Dynamics of  Straight Vortex}
It is simplest to consider a straight vortex~\cite{Svid00}, which applies
to a disk-shape condensate with $R_\perp\gg R_z$. Assume that the vortex is
located near the trap center at a transverse position $\vec{r}_0(t)$. In
this region, the trap potential does not change significantly on a scale of
order $\xi$.  The solution proceeds in two steps.

(a) First, consider the region near the vortex core that is assumed to move
with a transverse velocity $\vec{V}\perp \hat z$.  Transform to a co-moving
frame centered on the vortex core, where the trap potential exerts a force
proportional to $\vec{\nabla}_\perp V_{\rm tr}$ evaluated at
$\vec{r}_0(t)$.  The resulting  solution includes both the detailed
core structure and the ``asymptotic'' region $|\vec{r}_\perp -\vec{r}_{\perp
0}|\gg \xi$

(b)  Second, consider the region far from the vortex, where the core can be
treated as a distant singularity.  The short-distance behavior of this
solution includes the region  $ \xi \ll |\vec{r}_\perp -\vec{r}_{\perp
0}|$.  The two solutions must agree in the common region, which
determines the translational velocity $\vec{V}$ of the vortex line.

The details become intricate, but the final answer is elegant and physical:

\begin{equation}
\vec{V}=\frac{3\hbar}{4 M \mu}\left[\ln\left(\frac{R_\perp}{\xi}\right) -
\frac{2 M \Omega R_\perp^2}{3\hbar}
\right]\, \hat z\times\vec{\nabla}_\perp V_{\rm tr}\, ,\label{vel}
\end{equation}
where $R_\perp$ for an asymmetric trap is defined below Eq.~(\ref{approx}).
This expression has several important aspects

\begin{itemize}

\item The motion follows an equipotential line  along the
direction $\hat z\times\vec{\nabla}_\perp V_{\rm tr}$,
  conserving energy, as appropriate for the GP equation at zero
temperature.  For an asymmetric harmonic trap, the trajectory is elliptical.

\item For a nonrotating trap, the motion is counterclockwise, in the positive
sense.

\item With increasing external rotation, the translational velocity $\vec V$
decreases and vanishes at the special value $\Omega_m=
\frac{3}{2}(\hbar/MR_\perp^2)\ln(R_\perp/\xi)$ discussed below
Eq.~(\ref{approx}).

\item For $\Omega>\Omega_m$, the motion  as seen in the
rotating frame  is clockwise.

\item A detailed analysis of the normalization of
the  Bogoliubov amplitudes shows that the positive-norm state has the
frequency

\begin{equation}
\omega=
\frac{2\omega_x\omega_y}{\omega_x^2+\omega_y^2}\left(\Omega-\Omega_m\right).
\label{freq}
\end{equation}
This normal-mode frequency is negative (and hence locally unstable) for
$\Omega<\Omega_m$, but it becomes locally stable for $\Omega>\Omega_m$, in
agreement with the discussion below Eq.~(\ref{approx}).

\end{itemize}

\subsection{Inclusion of Curvature}

Equation (\ref{vel}) for the local velocity of a straight vortex oriented
along $\hat z$ can be generalized to include the possibility of a
different orientation of the  vector $\hat t$ locally tangent to the  vortex
core. In addition,   local curvature
$k$ of the vortex line defines a plane that includes both $\hat t$ and the
local  normal
$\hat n$,  inducing an additional
translational velocity  along the binormal vector $\hat b\equiv \hat t\times
\hat n$.   A detailed analysis~\cite{Svid00a} yields the translational
velocity of the element located at $\vec{r}_0$

\begin{equation}
\vec{V}(\vec{r}_0) = -\frac{\hbar}{2M}\left(\frac{\hat t\times
\vec{\nabla}V_{\rm tr}(\vec{r}_0)}{g|\Psi_{TF}|^2}+k\hat
b\right)\,\ln\left(\xi\sqrt{\frac{1}{R_\perp^2}+\frac{k^2}{8}}\,\right)
+\frac{2\vec{\nabla}V_{\rm
tr}(\vec{r}_0)\times\vec{\Omega}}{\nabla^2_\perp V_{\rm
tr}(\vec{r}_0)},\label{vela}
\end{equation}
where $\nabla^2_\perp$ is the Laplacian in the plane perpendicular to
$\vec{\Omega}$.  In the first term,  $|\Psi_{TF}|^2$ vanishes near the
  condensate boundary;  hence $\hat t\times \vec{\nabla}V_{\rm
tr}(\vec{r}_0)$ must also vanish there, implying that  the vortex is locally
perpendicular to the  surface.

Equation (\ref{vela}) allows a study of the dynamics of small-amplitude
displacements of the  vortex from the $z$ axis, when $x(z,t)$ and $y(z,t)$
obey coupled equations. In the limit $\omega_z=0$, there is no confinement
in the $z$ direction, and the density is independent of $z$.  The
resulting two-dimensional dynamics  exhibits helical solutions that are
linear combinations of two plane standing waves.

	More generally, for $\omega_z\neq 0$, the density near the 
$z$ axis has the
TF parabolic form, and the solutions become more complicated. It is
convenient to define the asymmetry parameters $\alpha = R_x^2/R_z^2$ and
$\beta = R_y^2/R_z^2$, where $\alpha> 1$, $\beta> 1$ indicate a disk
shape and $\alpha< 1$,
$\beta< 1$ indicate a cigar shape.

\begin{itemize}
\item For a nonrotating trap with the  special asymmetry values    $\alpha=
2/[n(n+1)]$
(here, $n$ a positive integer), the effects of the nonuniform trap potential
and the curvature  just balance, and the condensate has
stationary  solutions with the vortex at rest in the $x z$
plane.  A disk-shape trap  has no such states, and the first one
occurs for the spherical trap with $\alpha = 1$.  The next such state  occurs
for
$\alpha = \frac{1}{3}$, when the condensate  is significantly elongated.
Similar considerations for $\beta$ apply to stationary states in the $y z$
plane.

\item For other values of $\alpha $ and $\beta$,  solutions necessarily
involve  motion of the  vortex line relative to the stationary condensate.

\item  Analytical solutions can be found for an  extremely flat  disk
with $\alpha\gg 1$ and $\beta\gg 1$, reproducing the frequency
$\Omega_m$ found in Eq.~(\ref{approx}).

\item  For small deformations of a vortex line in an axisymmetric trap with
$\alpha=\beta$, a disk-shape or spherical condensate ($\alpha \ge 1$) has
only a single (unstable)  precessing  normal mode with a negative frequency
$\omega_a<0$.  In this case, an external rotation $\Omega \ge \Omega_m =
|\omega_a|$ stabilizes the vortex.  For  these geometries,
$\Omega_m$ is less than the thermodynamic critical value $\Omega_c$.
In a spherical condensate, the one anomalous mode $|\omega_a|$ agrees with
the observed vortex precession frequency seen in the JILA
experiments~\cite{Ande00,Fede00}.

\item In contrast, an axisymmetric cigar-shape condensate has additional
negative-frequency precessing modes, and $\Omega_m$ can exceed $\Omega_c$
for sufficiently elongated condensates; such behavior seems relevant
for the ENS experiments~\cite{Madi99,Chev00,Fede00}, where $\Omega_m\approx
1.7
\Omega_c$ is close to the observed rotation speed  for creating
  the first vortex.
 
\item For small axisymmetric deviations from a spherical trap, a straight
vortex line can execute large-amplitude periodic trajectories.  In this case,
the vortex line becomes invisible when it tips away from the line of sight,
and it then periodically returns to full visibility.  Such revivals agree
with preliminary observations at JILA~\cite{Corn00}.
 
\end{itemize}


\begin{thebibliography}{8.}
\addcontentsline{toc}{section}{References}


\bibitem{Ande95}  M.~H.~Anderson {\it et al.\/}, Science~{\bf 269}, 198
(1995).

\bibitem{Brad95}  C.~C.~Bradley {\it et al.\/},  Phys.~Rev.~Lett.~{\bf 75},
1687 (1995).

\bibitem{Davi95}  K.~B.~Davis {\it et al.\/}, Phys.~Rev.~Lett.~{\bf 75}, 3969
(1995).

\bibitem{Matt99}  M.~R.~Matthews {\it et al.\/}, Phys.~Rev.~Lett.~{\bf 83},
2498 (1999).

\bibitem{Madi99}  K.~W.~Madison {\it et al.\/},
Phys.~Rev.~Lett.~{\bf 84}, 806 (2000).

\bibitem{Madi00}  K.~W.~Madison {\it et al.\/},
e-print cond-mat/0004037.


\bibitem{Chev00}   F.~Chevy, K.~W.~Madison, and J.~Dalibard,
e-print cond-mat/0005221.

\bibitem{Ande00}  B.~P.~Anderson {\it et al.\/}, e-print cond-mat/0005368.


\bibitem{Gros61}  E.~P.~Gross, Nuovo Cimento {\bf 20}, 454 (1961).

\bibitem{Pita61}  L.~P.~Pitaevskii, Zh.~Eksp.~Teor.~Fiz.~{\bf 40}, 646
(1961) [Sov.~Phys. JETP {\bf 13}, 451 (1961)].

\bibitem{Dalf99}  F.~Dalfovo {\it et al.\/}, Rev.~Mod.~Phys.~{\bf 71}, 463
(1999).

\bibitem{Fett96} A.~L.~Fetter, Phys.~Rev.~A {\bf 53}, 4245 (1996).

\bibitem{Baym96} G.~Baym and C.~J.~Pethick, Phys.~Rev.~Lett.~{\bf 76}, 6
(1996).

\bibitem{Svid00}  A.~A.~Svidzinsky and A.~L.~Fetter, Phys.~Rev.~Lett.~{\bf
84}, 5919 (2000); e-print cond-mat/9811348.

\bibitem{Dalf97}  F.~Dalfovo {\it et al.\/}, Phys.~Rev.~A {\bf 56}, 3840
(1997).


\bibitem{Fede99}  D.~L.~Feder, C.~W.~Clark, and B.~I.~Schneider,
Phys.~Rev.~A {\bf 61}, 011601(R) (1999).

\bibitem{Isos99}  T.~Isoshima and K.~Machida, Phys.~Rev.~A {\bf 60}, 3313
(1999).

\bibitem{Fede00}  D.~L.~Feder {\it et al.}, e-print cond-mat/0009086.

\bibitem{Fett72}  A.~L.~Fetter, Ann.~Phys.~(N.Y.) {\bf 70}, 67 (1972).

\bibitem{Land41}  L.~D.~Landau, J.~Phys.~(USSR) {\bf 5}, 71 (1941).

\bibitem{Dodd97}  R.~J.~Dodd {\it et al.\/},
Phys.~Rev.~A {\bf 56}, 587 (1997).

\bibitem{Rokh97} D.~S.~Rokhsar, Phys.~Rev.~Lett.~{\bf 79}, 2164 (1997).

\bibitem{Stri96}  S.~Stringari, Phys.~Rev.~Lett.~{\bf 77}, 2360 (1996).

\bibitem{Svid98}   A.~A.~Svidzinsky and A.~L.~Fetter, Phys.~Rev.~A {\bf 58},
3168 (1998).

\bibitem{Zamb98}  F.~Zambelli and S.~Stringari, Phys.~Rev.~Lett.~{\bf 81},
1754 (1998).

\bibitem{Corn00}  E.~Cornell, private communication.

\bibitem{Pism91}  L.~M.~Pismen and J.~Rubinstein, Physica D {\bf 47}, 353
(1991).

\bibitem{Rubi94}  B.~Y.~Rubinstein and L.~M.~Pismen, Physics D {\bf 78}, 1
(1994).


\bibitem{Svid00a}   A.~A.~Svidzinsky and A.~L.~Fetter, e-print
cond-mat/0007139.



\end{thebibliography}
\end{document}